\documentstyle[12pt]{article}

\clearpage
\begin{document}

\hfill SOGANG-HEP 235/98


\vspace{0.5cm}

\begin{center}
{\large \bf Lagrangian Approach of the First Class Constrained Systems}
\end{center}

\vspace{0.5cm}

\begin{center}
Yong-Wan Kim$^{\dagger}$, Seung-Kook Kim$^{*}$
and Young-Jai Park$^{\dagger}$ \\
\end{center}
\begin{center}
{\it \ $^{\dagger}$ Department of Physics 
and Basic Science Research Institute \\
Sogang University, C.P.O. Box 1142, Seoul 100-611, Korea} \\
and\\
{\it \ $^{*}$ Department of Physics, Seonam University,
Namwon, Chonbuk 590-170, Korea}
\end{center}

\vspace{1cm}

\begin{center}
{\bf ABSTRACT}
\end{center}

We show how to systematically derive the exact form of 
local symmetries for the abelian Proca and CS models, 
which are converted into first class constrained systems 
by the BFT formalism, in the Lagrangian formalism. 
As results, without resorting to a Hamiltonian formulation
we obtain the well-known U(1) symmetry for the gauge 
invariant Proca model, while showing that for the CS model
there exist novel symmetries as well as the usual symmetry
transformations.

\vspace{1cm}



\newpage
\begin{center}
\section{\bf Introduction}
\end{center}

Field-Antifield formalism \cite{BV} is based on an analysis of local 
symmetries of a Lagrangian, and has a great adventage of representing 
manefestly covariant formulation of a theory. 
However, in general, local gauge symmetries are not systematically 
obtained in an action while constructing a Lagrangian.
Even though they are related with
the generalized Bianchi-like identities \cite{sudarshan},
it may be difficult to see the full local symmetries
for some complicated Lagrangians.
 
On the other hand, the Batalin, Fradkin, and Tyutin (BFT) 
Hamiltonian method \cite{BFT} has been applied 
to several second class constrained systems \cite{banerjee,KP,KPKK}, 
which yield the strongly involutive first class
constraint algebra in an extended phase space by introducing new fields. 
Recently, we have quantized other interesting models
including the Proca models by using our improved BFT formalism 
\cite{KKPPY,KPPY}. 
However, the Hamiltonian approach \cite{dirac,henneaux,FIK,BFT} 
to the quantization of constrained systems has the drawback of not 
necessarily leading to a manifestly Lorentz covariant partition function.  
This problem is also avoided in the Lagrangian field-antifield approach.
In this respect, the systematic and exhaustive determinations of local 
symmetries constitute an integral part of the field-antifield quantization 
program.

In this paper, we will consider Lagrangian approach 
of the first class constrained systems.
In section 2, after embedding the abelian Proca model on the extended 
phase space by the BFT method, we explicitly show how
to derive the exact form of the well-known local symmetry 
of the first class Proca model as a simple example 
from the view of Lagrangian
without resorting to a Hamiltonian formulation. 
In section 3, we apply the Lagrangian approach to abelian pure
Chern-Simons (CS) model which is invariant under the U(1) gauge 
transformation but has still second class constraint due to the symplectic 
structure. As results, we show that for the embedded symplectic structure
of the CS model there exist additional novel local symmetries 
as well as the usual U(1) gauge symmetry.
Our conclusions are given in section 4.

\vspace{1cm}
\begin{center}
\section {\bf Proca Model}
\end{center}

We first consider the abelian Proca model \cite{KPPY} 
whose dynamics are given by
\begin{equation}
\label{1}
S = \int d^4x \
             [ -\frac{1}{4} F_{\mu\nu} F^{\mu\nu}
             + \frac{1}{2} m^2 A_\mu A^\mu ],
\end{equation}
where $F_{\mu\nu}=\partial_\mu A_\nu-\partial_\nu A_\mu$, and
$g_{\mu\nu} = diag(+,-,-,-)$.
In the Hamiltonian formulation the canonical momenta of gauge fields are 
given by $\pi_{0} = 0$, and $\pi_{i} = F_{i0}$.
Then, $\Omega_1 = \pi_0 \approx 0$ is a primary constraint, 
and the canonical Hamiltonian 
\begin{equation}
\label{2}
H_c = \int d^3x \left[
                \frac{1}{2} \pi_i^2 + \frac{1}{4} F_{ij} F^{ij}
              + \frac{1}{2} m^2 \{ (A^0)^2 + (A^i)^2 \}
              - A_0\Omega_2  
              \right].
\end{equation}
Here $\Omega_2$ is the Gauss' law constraint
which comes from the time evolution of $\Omega_1$ with the primary Hamiltonian
$H_p=H_c+\int d^3x \ v^1 \Omega_1$ as 
\begin{equation}
\label{3}
\Omega_2 = \partial^i \pi_i + m^2 A^0 \approx 0.
\end{equation}

We now convert these second class constraints into the corresponding
first class ones via {\it a la} BFT Hamiltonian embedding. 
This BFT method is by now well-known, and thus we would like to avoid 
the explicit calculation here and quote the results of Ref. \cite{KPPY}.
The effective first class constraints $\widetilde{\Omega}_{i}$ 
are given by 
\begin{eqnarray}
\label{4}
\widetilde{ \Omega}_{i}&=&\Omega_{i}+\Omega^{(1)}_{i}  
=\Omega_{i} +m \Phi^{i}.
\end{eqnarray}
with the introduction of auxiliary fields $\Phi^i$ satisfying
$\{\Phi^i(x),\Phi^j(y)\}=\omega^{ij}(x,y)=\epsilon^{ij} 
\delta^3(x-y)$, 
and the first class Hamiltonian $\widetilde{H}$
corresponding to the canonical Hamiltonian $H_{c}$ by 
\begin{eqnarray}
\label{5}
\widetilde{H}(A^{\mu}, \pi_{\nu}; \Phi^{i} )
    =    H_{c}(A^{\mu}, \pi_{\nu})+
              \int d^3x \left[
                 \frac{1}{2} (\Phi^{2})^{2}
                 + \frac{1}{2}( \partial_{i} \Phi^{1} )^{2}
                 + m \Phi^{1} \partial_i A^i
                 - \frac{1}{m} \Phi^{2}
                     \widetilde{\Omega}_{2}
                                \right], 
\end{eqnarray}
which is  strongly involutive, i.e.,
$\{ \widetilde{\Omega}_{i}, \widetilde{H} \} =0$.

It seems appropriate to comment on generators of local symmetry transformation
in the Hamiltonian formulation, which are given by the first class constraints. 
Defining the generators by
\begin{equation}
\label{6}
G=\sum^2_{\alpha=1}\int d^2x~(-1)^{\alpha+1}\epsilon^\alpha(x)\Omega_\alpha(x),
\end{equation}
we have 
\begin{eqnarray}
\label{7}
&&\delta A^0=\epsilon^1,~~~~~~\delta \pi_0=m^2\epsilon^2, \nonumber\\
&&\delta A^i=\partial^i\epsilon^2,~~~~\delta \pi_i=0,\nonumber\\
&&\delta\rho=-\epsilon^2,~~~~~~\delta \pi_\rho=-m^2\epsilon^1.
\end{eqnarray}
Here we inserted $(-1)^{\alpha+1}$ factor in Eq. (\ref{6}) 
in order to make the gauge transformation usual, 
and also identified the new variables
$\Phi^i$ as a canonically conjugate pair
($\rho$, $\pi_\rho$) in the Hamiltonian formalism through
$(\Phi^i) \to  (m \rho, \frac{1}{m}\pi_\rho)$.
Now, it can be easily seen that the extended action
\begin{equation}
\label{8}
S_{\rm E}=\int d^4x (\pi_\mu\dot{A}^\mu+\pi_\rho\dot{\rho}-\widetilde{H})
\end{equation}
is invariant under the gauge transformations (\ref{7}) 
with $\epsilon^1=\partial^0\epsilon^2$.

Next, from the partition function given by the Faddeev-Popov
formula \cite{faddeev} as 
\begin{equation}
\label{9}
Z=  \int  {\cal D} A^\mu
          {\cal D} \pi_\mu
          {\cal D} \rho
          {\cal D} \pi_\rho
               \prod_{i,j = 1}^{2} \delta(\widetilde{\Omega}_i)
                           \delta(\Gamma_j)
               \mbox{det} \mid \{ \widetilde{\Omega}_i, \Gamma_j \} \mid
                e^{iS},
\end{equation}
where
\begin{equation}
\label{10}
S  =  \int d^4x \left(
         \pi_\mu {\dot A}^\mu + \pi_\rho {\dot \rho} - \widetilde{\cal H}
                \right),
\end{equation}
and $\Gamma_j$ are the gauge fixing functions.
One could integrate all the momenta out with the delta functional in Eq.
(\ref{9}). As results, we have the well-known action 
\begin{equation}
\label{11}
S = \int d^4x {\cal L} = \int d^4x  \left[ - \frac{1}{4}  F_{\mu\nu} F^{\mu\nu}
                 + \frac{1}{2} m^2 (A_\mu + \partial_\mu \rho)^2\right],
\end{equation}
which is invariant under $\delta A^\mu=\partial^\mu\epsilon^2$
and $\delta\rho=-\epsilon^2$.

Now, we are ready to use a recently proposed
Lagrangian approach \cite{shirzad} of constrained systems 
in the configuration space.
Starting from the gauge invariant action (\ref{11}),
our purpose is to find the gauge transformation rules systematically
without resorting to the language of the Hamiltonian formulation.

The equations of motion following from (\ref{11}) are of the form
\begin{equation}
\label{12}
L^{(0)}_i(x)=\int d^3y \left[W^{(0)}_{ij}(x,y)\ddot\varphi^j(y)
+\alpha^{(0)}_i(y)\delta^3(x-y)\right]=0, \ i=1,2,...,5,
\end{equation}
where
$W^{(0)}_{ij}(x,y)$ is the Hessian matrix
\begin{eqnarray}
\label{13}
W^{(0)}_{ij}(x,y)&:=&\frac{\delta^2{\cal L}}{\delta\dot\varphi^i(x)\delta\dot
\varphi^j(y)}\nonumber\\
&=& 
\left(\begin{array}{ccccc}
       0 & 0 & 0 & 0 & 0 \\
       0 & 1 & 0 & 0 & 0 \\
       0 & 0 & 1 & 0 & 0 \\
       0 & 0 & 0 & 1 & 0 \\
       0 & 0 & 0 & 0 & m^2
       \end{array} 
       \right)\delta^3(x-y)
=\widetilde{W}^{(0)}_{ij}\delta^3(x-y),\\
\label{14}
(\varphi^i)^{\rm T}(x)
&=&(A^0,A^1,A^2,A^3,\rho)(x),\\
\label{15}
(\alpha^{(0)}_i)^{\rm T}(x)&:=&\int d^3y\left[
\frac{\partial^2{\cal L}}{\partial\varphi^j(y)\partial\dot\varphi^i(x)}
\partial\dot\varphi^j(y)\right]-\frac{\partial {\cal L}}{\partial\varphi^i(x)}
\nonumber\\
&=&(\alpha_{A^0},\alpha_{A^1},\alpha_{A^2},\alpha_{A^3},
\alpha_{\rho})(x)
\end{eqnarray}
with
\begin{eqnarray}
\label{16}
&&\alpha_{A^0}=\partial_i(\dot{A}^i+\partial_iA^0)-m^2(A^0+\dot\rho),
\nonumber\\
&&\alpha_{A^i}=\partial_i\dot{A}^0-\partial_jF^{ij}+m^2(A^i-\partial_i\rho),
\nonumber\\
&&\alpha_{\rho}=m^2\dot{A}^0+m^2\partial_i(A^i-\partial_i\rho).
\end{eqnarray}
Since the Hessian matrix (\ref{13}) is of rank four,
there exists a ``zeroth generation'' null eigenvector
$\lambda^{(0)}(x,y)$ satisfying
\begin{equation}
\label{17}
\int d^3y\ \lambda^{(0)}_i(x,y)\ W^{(0)}_{ij}(y,z)=0.
\end{equation}
For simplicity, let us normalize it to have components
\begin{equation}
\label{18}
\lambda^{(0)}(x,y)=(1,0,0,0,0)\delta^3(x-y).
\end{equation}
Correspondingly we have a ``zeroth generation" constraint
in the Lagrangian sense as
\begin{equation}
\label{19}
\Omega^{(0)}_1(x)=\int d^3y \ \lambda^{(0)}_i(x,y) L^{(0)}_i(y)=L^{(0)}_1(x)
=\alpha_{A^0}=0,
\end{equation}
when multiplied with the equations of motion (\ref{12}).

We now require the ``zeroth generation'' Lagrange constraint (\ref{19})
to vanish in the evolution of time, and add this to the equations
of motion (\ref{12}) through the equation of $\dot{\Omega}^{(0)}_1=0$. 
The resulting set of six equations may be summarized
in the form of the set of ``first generation'' equations,
$L^{(1)}_{i_1}(x)=0,\ i_1=1,...,6$, with
\begin{equation}
\label{20}
L_{i_1}^{(1)}(x)=\left\{\begin{array}{ll}
L_i^{(0)},\ i=1,...5,\\
\frac{d}{dt}(\lambda^{(0)}_iL^{(0)}_i).
\end{array}\right.
\end{equation}
$L_{i_1}^{(1)}(x)$  can be written in the general form
\begin{equation}
\label{21}
L_{i_1}^{(1)}(x)
=\int d^3y \left[W_{i_1j}^{(1)}(x,y)\ddot\varphi^j(y)
            +\alpha^{(1)}_{i_1}(y)\delta^3(x-y)\right]=0, ~~~i_1=1,\cdot,6,
\end{equation}
where
\begin{equation}
\label{22}
W^{(1)}_{i_1j}(x,y)=\left(\begin{array}{cc}
\\
\widetilde{W}^{(0)}_{ij}\\
\\
\hline\\
\begin{array}{ccccc}
0&-\partial^1_x&-\partial^2_x&-\partial^3_x&-m^2
\end{array}
\end{array}
\right)\delta^3(x-y),
\end{equation}
and
\begin{equation}
\label{23}
(\alpha^{(1)}_{i_1})^{\rm T}(x)=( (\alpha^{(0)}_i)^{\rm T}, \alpha^{(1)}_6)(x)
\end{equation}
with
\begin{eqnarray}
\label{24}
\alpha^{(1)}_6=\partial_i\partial_i\dot{A}^0-m^2\dot{A}^0.
\end{eqnarray}

We now repeat the previous analysis taking Eq. (\ref{21}) as a
starting point, and looking for solutions of a first generation 
null eigenvector as
\begin{equation}
\label{25}
\int d^3y \ \lambda^{(1)}_{i_1}(x,y)W^{(1)}_{i_1j}(y,z)=0.
\end{equation}
Since $W^{(1)}_{i_1j}(x,y)$ is still of rank four, there exit 
two null eigenvectors, $\lambda^{(1)}$ and $\Sigma^{(1)}$. 
The $\lambda^{(1)}$ is the previous null
eigenvector (\ref{18}) with an extension such as 
$\lambda^{(1)}(x,y)=(\lambda^{(0)}_i,0)$, and   
the other null eigenvector $\Sigma^{(1)}$ of $W^{(1)}_{i_ij}(x,y)$ is of the
form $(0,\partial^1_x,\partial^2_x,\partial^3_x,1,1)v(x)
\delta^3(x-y)$. We could thus choose it as
\begin{equation}
\label{26}
\Sigma^{(1)}_{i_1}(x,y)=(0,\partial^1_x,\partial^2_x,\partial^3_x,1,1)
\delta^3(x-y).
\end{equation}
The associated constraint is found to vanish ``identically''
\begin{equation}
\label{27}
\Omega^{(1)}_2(x)=\int d^3y \ \Sigma^{(1)}_{i_1}(x,y)L^{(1)}_{i_1}(y)
=\partial^i \alpha^{(1)}_{A^i} + \alpha^{(1)}_5 
+ \alpha^{(1)}_6=0.
\end{equation}
The algorithm ends at this point.

The local symmetries of the action (\ref{11}) 
are encoded in the identity (\ref{27}). 
Recalling (\ref{12}) and (\ref{21})
we see that the identity (\ref{27}) can be rewritten as follows
\begin{eqnarray}
\label{28}
\Omega^{(1)}_2(x)=\partial^iL^{(0)}_i+L^{(0)}_5+\frac{d}{dt}L^{(0)}_1=0.
\end{eqnarray}
This result is a special case of a general theorem stating
\cite{shirzad} that the identities $\Omega^{(l)}_k\equiv 0$ 
can always be written in the form
\begin{equation}
\label{29}
\Omega^{(l)}_k=\sum_{s=0} \int d^3y \left((-1)^{s+1}
                \frac{d^s}{dt^s}\phi^{i(s)}_k(x,y)L^{(0)}_i(y)\right).
\end{equation}
For the Proca case we have the following relations
\begin{eqnarray}
\label{30}
&&\phi^{2(0)}_2(x,y)=-\partial^1\delta^3(x-y),\nonumber\\
&&\phi^{3(0)}_2(x,y)=-\partial^2\delta^3(x-y),\nonumber\\
&&\phi^{4(0)}_2(x,y)=-\partial^3\delta^3(x-y),\nonumber\\
&&\phi^{5(0)}_2(x,y)=-\delta^3(x-y),\nonumber\\
&&\phi^{1(1)}_2(x,y)=\delta^3(x-y).
\end{eqnarray}
It again follows from general considerations \cite{shirzad}
that the action (\ref{11}) is invariant under the transformation
\begin{equation}
\label{31}
\delta\varphi^i(y)=\sum_{k}\int d^3x \ \left(\Lambda_k(x)\phi^{i(0)}_k(x,y)
                    + \dot{\Lambda}_k(x)\phi^{i(1)}_k(x,y)\right).
\end{equation}
For the case in question this corresponds to the transformations
\begin{eqnarray}
\label{32}
&&\delta A^\mu(x)=\partial^\mu\Lambda_2,\nonumber\\
&&\delta \rho(x)=-\Lambda_2.
\end{eqnarray}
These are the set of symmetry transformations which is identical with 
the previous result (\ref{7}) of the extended Hamiltonian formalism,
when we set $\epsilon^1=\partial^0\epsilon^2$ and $\epsilon^2=\Lambda_2$, 
similar to the Maxwell case \cite{comment}.
As results, we have systematically derived the set of symmetry 
transformations starting from the Lagrangian of the first class Proca model.

\vspace{1cm}
\begin{center}
\section{\bf Chern-Simons Model}
\end{center}

Similar to the Proca case, we have recently applied the BFT method 
to the pure CS theory whose dynamics are given by
\begin{equation}
\label{33}
S=\int d^3x \ \frac{\kappa}{2}\epsilon_{\mu\nu\rho}A^\mu\partial^\nu A^\rho.
\end{equation}
This is invariant under the U(1) gauge transformation, $\delta A^\mu
=\partial^\nu\Lambda$, but has still second class constraints due to the
symplectic structure of the CS theory.
As a result, we have obtained the following fully first class CS action 
\cite{KPKK} as
\begin{equation}
\label{34}
S=\int d^3x \left( \frac{\kappa}{2} \epsilon_{\mu \nu \rho} A^{\mu} 
\partial^{\nu} A^{\rho} - \frac{1}{2} \epsilon_{ij} \Phi^i \dot \Phi^j 
+ \sqrt \kappa \Phi^j F_{0j} \right),
\end{equation}
where $\Phi^i$ satisfy the relation $\{\Phi^i(x),\Phi^j(y)\}=\epsilon^{ij}
\delta(x-y)$.
Then, we may raise a question what is the symmetry transformation
corresponding to the additional first class constraints originated from
the symplectic structure of the CS theory. 
We would like to find them through the similar 
analysis of the previous Lagrangian approach.

The equations of motion following from (\ref{34}) are of the form
\begin{eqnarray}
\label{35}
(L_i^{(0)})^T (x) &=& ( L_{A^0}, L_{A^1}, L_{A^2}, L_{\Phi^1}, L_{\Phi^2}),\\
\label{36}
(\alpha_i^{(0)})^T (x) &=& ( \alpha_{A^0}, \alpha_{A^1}, \alpha_{A^2},
\alpha_{\Phi^1}, \alpha_{\Phi^2} ).
\end{eqnarray}
The starting Hessian matrix is trivial for this pure CS case
due to the first order Lagrangian as follows
\begin{equation}
\label{37}
W^{(0)}_{ij}(x,y) = \left(\begin{array}{ccccc}
                    0 & 0 & 0 & 0 & 0 \\
                    0 & 0 & 0 & 0 & 0 \\
                    0 & 0 & 0 & 0 & 0 \\
                    0 & 0 & 0 & 0 & 0 \\
                      0 & 0 & 0 & 0 & 0
                  \end{array} 
                   \right)\delta^2(x-y)
                 \equiv \widetilde{W}^{(0)}_{ij}\delta^2(x-y),
\end{equation}
which shows that there are no true dynamical degrees of freedom.
Explicitly, the equations of motion are given by
\begin{eqnarray}
\label{38}
L_{A^0} &=& -\kappa\epsilon_{ij}\partial^iA^j - {\sqrt\kappa}\partial_1
\Phi^1 - {\sqrt \kappa} \partial_2 \Phi^2 = \alpha_{A^0}, \nonumber \\
L_{A^1} &=& \kappa \dot A^2 - {\sqrt \kappa} \dot \Phi^1 + \kappa 
\partial_2 A^0 = \alpha_{A^1}, \nonumber \\
L_{A^2} &=& - \kappa \dot A^1 - {\sqrt \kappa} \dot \Phi^2 - \kappa 
\partial_1 A^0 = \alpha_{A^2}, \nonumber \\
L_{\Phi^1} &=& \dot \Phi^2 + {\sqrt \kappa} ( \dot A^1 + \partial_1 A^0 )
 = \alpha_{\Phi^1}, \nonumber \\
L_{\Phi^2} &=& - \dot \Phi^1 + {\sqrt \kappa} (\dot A^2 + \partial_2 A^0 )
= \alpha_{\Phi^2}.
\end{eqnarray}
Since the Hessian matrix (\ref{37}) is of rank zero, 
there exist five ``zeroth generation'' null eigenvectors as
\begin{equation}
\label{39}
\lambda^{(0)a}_i (x,y) = \delta^a_i \delta^2(x-y), ~~~~i, a= 1, \cdots, 5.
\end{equation}
Correspondingly we have  ``zeroth generation'' Lagrange constraints
\begin{equation}
\label{40}
\Omega^{(0)}_i = L^{(0)}_i = \alpha^{(0)}_i.
\end{equation}
Moreover, the equations of motion (\ref{38}) are not independent. We can
thus obtain the following identical relations as
\begin{eqnarray}
\label{41}
&& \alpha^{(0)}_2 - {\sqrt \kappa} \alpha^{(0)}_5 = 0, \nonumber \\
&& \alpha^{(0)}_3 + {\sqrt \kappa} \alpha^{(0)}_4 = 0.
\end{eqnarray}
As results, we can rewrite the ``zeroth generation'' Lagrange constraints
as follows 
\begin{equation}
\label{42}
\Omega^{(0)}_i = \left\{ \begin{array}{ll}
                         \Omega^{(0)}_{\bar i} = \Omega^{(0)}_i,
                          ~~~~ i,\bar i = 1,2,3,\\
                          \Omega^{(0)}_{\hat i},~~~~~~~~~~~~~\hat i = 1,2,
                            \end{array} \right.
\end{equation}
where the bar in the subscript denotes the independent constraints
while the carrot identities as
\begin{eqnarray}
\label{43}
\Omega^{(0)}_{\hat 1} &=& \alpha^{(0)}_2 - {\sqrt \kappa} \alpha^{(0)}_5
= L^{(0)}_2 - {\sqrt \kappa} L^{(0)}_5 = 0, \nonumber \\
\Omega^{(0)}_{\hat 2} &=& \alpha^{(0)}_3 + {\sqrt \kappa} \alpha^{(0)}_4
= L^{(0)}_3 + {\sqrt \kappa} L^{(0)}_4 = 0.
\end{eqnarray} 
We now require the independent ``zeroth generation'' Lagrange constraints
to vanish in time evolution.
Then, the resulting set of eight equations may be summarized 
in the form of the set of ``first generation'' equations, 
$L_{i_1}^{(1)}=0,~i_1 =1, \cdots , 8 $, with
\begin{equation}
\label{44}
L_{i_1}^{(1)}(x)=\left\{\begin{array}{ll}
L_i^{(0)},\ ~~~~i=1,...5,\\
\frac{d}{dt}(\Omega^{(0)}_{\bar i}).
\end{array}\right.
\end{equation}
$L_{i_1}^{(1)}(x)$ can be written in the general form as
\begin{equation}
\label{45}
L_{i_1}^{(1)}(x)
=\int d^2y \left[W_{i_1j}^{(1)}(x,y)\ddot\varphi^j(y)
            +\alpha^{(1)}_{i_1}(y)\delta^2(x-y)\right]=0,
\end{equation}
where the Hessian matrix is given by
\begin{equation}
\label{46}
W^{(1)}_{i_1j}(x,y)=\left(\begin{array}{cc}
\\
\widetilde{W}^{(0)}_{ij}\\
\\
\hline\\
\begin{array}{ccccc}
0& 0& 0& 0& 0 \\
0& 0& \kappa & - {\sqrt \kappa} & 0 \\
0& -\kappa &0 &0 & -{\sqrt \kappa}
\end{array}
\end{array}
\right)\delta^2(x-y),
\end{equation}
and
\begin{equation}
\label{47}
(\alpha_{i_1}^{(1)} )^T (x) = ((\alpha_i^{(0)})^T, \alpha_6^{(1)},
\alpha_7^{(1)}, \alpha_8^{(1)}),
\end{equation}
with
\begin{eqnarray}
\label{48}
\alpha_6^{(1)} &=& - \kappa \epsilon_{ij} \partial^i \dot A^j 
- \kappa \partial_1 \dot\Phi^1 - {\sqrt\kappa} \partial_2 \dot \Phi^2,
\nonumber \\
\alpha_7^{(1)} &=& \kappa\partial_2 \dot A^0, \nonumber \\
\alpha_8^{(1)} &=& -\kappa\partial_1 \dot A^0.
\end{eqnarray}

We now repeat the previous analysis taking Eq. (\ref{45}) as a
starting point, 
and looking for solutions of a first generation null eigenvector as
\begin{equation}
\label{49}
\int d^2y \ \lambda^{(1)}_{i_1}(x,y)W^{(1)}_{i_1j}(y,z)=0.
\end{equation}
Since $W^{(1)}_{i_1j}(x,y)$ is of rank two, there exist 
six null eigenvectors, $\lambda^{(1)a}_i$ and $\Sigma^{(1)}$. 
These $\lambda^{(1)a}_i$ are the previous null
eigenvector (\ref{39}) with an extension such as 
$\lambda^{(1)a}_{i_1}(x,y) = (\lambda^{(0)a}_i,0)$, and
a new null eigenvector $\Sigma^{(1)}$ as
\begin{equation}
\label{50}
\Sigma^{(1)}(x,y) = ( 0,0,0,0,0,1,0,0) \delta^2(x-y).
\end{equation}
The associated constraint of $\Sigma^{(1)}(x,y)$ generates one more
Lagrange constraint as
\begin{equation}
\label{51}
\Omega^{(1)}_{\bar 1} (x,y) 
= \int d^2 y \Sigma^{(1)}(x,y) L^{(1)}_{i_1}
(y) = - \kappa \epsilon_{ij} \partial^i \dot A^j 
- \kappa \partial_1 \dot \Phi^1 - {\sqrt \kappa} \partial_2 \dot \Phi^2 =0.
\end{equation}
Now, the resulting set of nine equations may be summarized 
in the form of  ``second generation'', 
$L_{i_2}^{(2)}=0,~i_1 =1, \cdots , 9 $, with
\begin{equation}
\label{52}
L_{i_2}^{(2)}(x)=\left\{\begin{array}{ll}
L_{i_1}^{(1)},\ \ \ i_1 =1,\cdots,8, \\
\frac{d}{dt}(\Omega^{(1)}_{\bar 1}),
\end{array}\right.
\end{equation}
which can be written in the general form
\begin{equation}
\label{53}
L_{i_2}^{(2)}(x)
=\int d^2y \left[W_{i_2j}^{(2)}(x,y)\ddot\varphi^j(y)
            +\alpha^{(2)}_{i_2}(y)\delta^2(x-y)\right]=0,
\end{equation}
where
\begin{equation}
\label{54}
W^{(2)}_{i_2j}(x,y)=\left(\begin{array}{cc}
\\
\widetilde{W}^{(0)}_{ij}\\
\\
\hline\\
\begin{array}{ccccc}
0& 0& 0& 0& 0 \\
0& 0& \kappa & - {\sqrt \kappa} & 0 \\
0& -\kappa &0 &0 & -{\sqrt \kappa} \\
0& \kappa \partial^2_x & -\kappa \partial^1_x & {\sqrt \kappa} \partial^1_x
& {\sqrt \kappa} \partial^2_x
\end{array}
\end{array}
\right)\delta^3(x-y),
\end{equation}
and
\begin{equation}
\label{55}
(\alpha_{i_2}^{(2)} )^T (x) = ((\alpha_{i_1}^{(1)})^T, \alpha_9^{(2)}).
\end{equation}
with identically vanishing component of $\alpha_9^{(2)}=0$.

We now repeat the analysis starting from Eq. (\ref{53}).
Since $W^{(2)}_{i_2j}(x,y)$ is still of rank two, 
there exists one more new null eigenvector $\Sigma^{(2)}$ as
\begin{equation}
\label{56}
\Sigma^{(2)}(x,y) = (0,0,0,0,0,0,\partial^1_x,\partial^2_x, 1) 
\delta^2(x-y).
\end{equation}
with the properly extended previous null eigenvectors $\lambda^{(1)a}_{i_1}$
and $\Sigma^{(1)}$. 
The associated constraint is now found to vanish ``identically''
\begin{equation}
\label{57}
\Omega^{(2)}_{\hat 1}(x) = \int d^3y \lambda^{(2)}_{i_2} (x,y) L_{i_2}(y)
= \partial^1 L_7(x) + \partial^2 L_8(x) + L_9(x) = 0,
\end{equation}
and the algorithm stops at this stage.
As results, we gather all the identities, which will provide the explicit
form of symmetry transformations, as follows
\begin{eqnarray}
\label{58}
\Omega^{(0)}_{\hat 1} &=& \Omega^{(0)}_1
= L_2 - {\sqrt \kappa} L_5 =0, \nonumber \\
\Omega^{(0)}_{\hat 2} &=& \Omega^{(0)}_2 
= L_3 + {\sqrt \kappa} L_4 =0, \nonumber \\
\Omega^{(2)}_{\hat 1} &=& \Omega^{(2)}_3 
= \partial^1 L_7 + \partial^2 L_8 + L_9 = \frac{d}{dt}(\partial^1 L_2 
+ \partial^2 L_3) + \frac{d^2}{dt^2} L_1 =0.
\end{eqnarray} 
Comparing these with Eq. (\ref{29}),
we have the following relations
\begin{eqnarray}
\label{59}
\phi_1^{2(0)} (x,y) &=& -\delta^2(x-y), \nonumber \\
\phi_1^{5(0)} (x,y) &=& {\sqrt \kappa} \delta^2(x-y), \nonumber \\
\phi_2^{3(0)} (x,y) &=& -\delta^2(x-y), \nonumber \\
\phi_2^{4(0)} (x,y) &=& -{\sqrt \kappa} \delta^2(x-y), \nonumber \\
\phi_3^{2(1)} (x,y) &=& -\partial^1_x \delta^2(x-y), \nonumber \\
\phi_3^{3(1)} (x,y) &=& -\partial^2_x \delta^2(x-y), \nonumber \\
\phi_3^{1(2)} (x,y) &=& -\delta^2(x-y),
\end{eqnarray}
and making use of the following general expression of
\begin{equation}
\label{60}
\delta\varphi^i(y)=\sum_{k}\int d^2x \ \left(\Lambda_k(x)\phi^{i(0)}_k(x,y)
                    + \dot{\Lambda}_k(x)\phi^{i(1)}_k(x,y) 
                    + \ddot{\Lambda}_k(x)\phi^{i(2)}_k(x,y)\right),
\end{equation}
we finally obtain the extended symmetry transformations of the first
class pure CS theory as
\begin{eqnarray}
\label{61}
\delta A^0 (x) &=&  \partial^0\dot\Lambda_3,\nonumber \\
\delta A^i (x) &=&  \partial^i \dot\Lambda_3+\Lambda_i, \nonumber \\
\delta \Phi^i (x) &=& \sqrt \kappa \epsilon^{ij} \Lambda_j.
\end{eqnarray}
Therefore, we see that the gauge parameter $\Lambda_3$ is related with
the usual U(1) gauge transformation, while $\Lambda_i~(i=1,2)$ 
generate novel symmetries originated from the symplectic structure 
of the CS theory. 
Note that if we further restrict the transformation as
$\dot\Lambda_3=\Lambda$, and $\Lambda_i=0$, 
then we easily recognize this novel extended symmetries reduce to the
original well-known U(1) symmetry.

\vspace{1cm}
\begin{center}
\section{\bf Conclusion}
\end{center}

In this paper we have considered the Lagrangian approach of
the first class abelian Proca and CS models. 
First, we have turned the second class Lagrangians into  
first class ones following the BFT method. 
Although the gauge invariant Lagrangian for the simple Proca model
corresponding to the first class 
Hamiltonian exhibits the well-known local symmetry,
we have explicitly shown, following the version of 
the Lagrangian approach, 
how this symmetry could be systematically derived on a
purely Lagrangian level, without resorting to a Hamiltonian formulation.
On the other hand, we have also studied the fully first class CS model
by embedding the so called symplectic structure on the extended space.
As results, we have found that there exist novel symmetries
as well as  the usual U(1) gauge symmetry by the Lagrangian method.
We hope that the Lagrangian approach employed in these derivation 
will be of much interest in the context of the field-antifield formalism.

\vspace{1cm}
\begin{center}
\section*{Acknowledgements}
\end{center}

The present study was supported by
the Basic Science Research Institute Program,
Ministry of Education, Project No. BSRI--98--2414.

\end{document}